\documentclass[11pt,a4paper]{article}
\usepackage{jheppub}
\usepackage{graphicx}  
\usepackage{dcolumn}   
\usepackage{bm}        
\usepackage{amssymb}   
\usepackage[latin1]{inputenc}
\usepackage{amsmath}
\usepackage{amsfonts}
\usepackage{amssymb}
\usepackage{setspace}
\usepackage{mathtools}
\usepackage{pstricks}
\usepackage{color}

\providecommand{\f}[2]{\frac{{#1}}{{#2}}}
\newcommand{\da}{\ensuremath{\dot{a}}}
\newcommand{\dda}{\ensuremath{\ddot{a}}}
\newcommand{\ddda}{\ensuremath{\dddot{a}}}
\newcommand{\ee}[1]{\begin{equation}#1\end{equation}}
\newcommand{\ea}[1]{\begin{align}#1\end{align}}

\title{Renormalization of the inflationary perturbations revisited}

\author[a]{Tommi Markkanen}
\affiliation[a]{Department of Physics, Imperial College London, SW7 2AZ, UK}

\abstract{In this work we clarify aspects of renormalization on curved backgrounds focussing on the potential ramifications on the amplitude of inflationary perturbations. We provide an alternate view of the often used adiabatic prescription by deriving a correspondence between the adiabatic subtraction terms and traditional renormalization. Specifically, we show how adiabatic subtraction can be expressed as a set of counter terms that are introduced by redefining the bare parameters of the action. Our representation of adiabatic subtraction then allows us to easily find other renormalization prescriptions differing only in the finite parts of the counter terms. As our main result, we present for quadratic inflation how one may consistently express the renormalization of the spectrum of perturbations from inflation as a redefinition of the bare cosmological constant and Planck mass such that the observable predictions coincide with the unrenormalized result.}

\emailAdd{t.markkanen@imperial.ac.uk}
\begin{document}
\begin{flushleft}
	\hfill		 IMPERIAL/TP/2017/TM/02
\end{flushleft}

\maketitle
\section{Introduction}
The earliest inflationary models were proposed over 30 years ago \cite{Guth:1980zm,Linde:1981mu,Albrecht:1982wi,Starobinsky:1979ty,Starobinsky:1980te} and to this day the framework of cosmological inflation is perhaps the most successful proposal for explaining several non-trivial features of the observed Universe \cite{Ade:2015lrj}.

In the simplest realisation of inflation an epoch of quasi-exponential expansion of space results from a classical field $\varphi$ slowly rolling down its potential during which its quantum fluctuations $\hat{\phi}$ get stretched to cosmological scales providing the necessary inhomogeneities that seed the formation of structure in the Universe. The inflationary perturbations are quantified by the spectrum of the co-moving curvature perturbation \cite{Sasaki:1986hm,Mukhanov:1988jd}, $\mathcal{R}$, which in the spatially flat gauge is sourced by fluctuations from the scalar field as
\ee{\mathcal{P}_{\mathcal{R}}=\f{H}{\dot{\varphi}}\mathcal{P}_{\phi}\,,\label{eq:R}}
where the power spectrum of a generic quantized variable is defined by a momentum space integral
\ee{\langle\hat{f}\,^2\rangle\equiv \int _0^\infty \f{dk}{k}\,\mathcal{P}_{f}\,,\label{eq:PS}}
with $|\mathbf{k}|\equiv k$. The current state-of-the-art measurements give for the observable long wavelength part of the spectrum \cite{Ade:2015lrj}
\ee{\mathcal{P}_{\mathcal{R}}=2.2\times 10^{-9}\,.}

Although rarely discussed in works focussing on Early Universe phenomenology, according to general field theory principles the spectrum of perturbations $\mathcal{P}_{\mathcal{R}}$ can potentially be modified by the renormalization counter terms one must unavoidably introduce to render the theory finite. For example, in general the variance of the field fluctuations $\langle\hat{\phi}^2\rangle$ is a divergent quantity and any physical result containing it will also have to contain counter terms, $\delta\phi^2$, removing the divergences. By way of (\ref{eq:PS}) a renormalized variance provides a definition for the renormalization of the power spectrum $\hat{\phi}$ 
\ee{\langle\hat{\phi}^2\rangle_R\equiv\langle\hat{\phi}^2\rangle-\delta\phi^2\equiv\int _0^\infty \f{dk}{k}\,\big(\mathcal{P}_{\phi}-\delta\mathcal{P}_{\phi}\big)\equiv \int _0^\infty \f{dk}{k}\,(\mathcal{P}_{\phi})_R\,,\label{eq:Rp0}}
and trivially via (\ref{eq:R}) for $(\mathcal{P}_\mathcal{R})_R\equiv\mathcal{P}_\mathcal{R}-\delta\mathcal{P}_\mathcal{R}$.
As the above shows, in general renomalization can have an effect on the amplitude of the inflationary perturbations, it is however widely assumed that renormalization may be performed in such a way that only the very ultraviolet contribution of the spectrum is modified with no effect on the observable long wavelength modes and robust inflationary predictions can be made without addressing proper renormalization of the theory. 

However, it has also been claimed that renormalization will unavoidably introduce a significant change on the amplitude of the spectrum perturbations from inflation, including  the long wavelength modes thus resulting in completely different predictions when compared with standard results \cite{Parker:2007}. The results of \cite{Parker:2007} have been the subject of a considerable amount of debate \cite{Agullo:2008ka, Agullo:2009vq, Agullo:2009zza, Agullo:2009zi, Agullo:2010ui, Agullo:2010hg, Agullo:2011qg, Bastero-Gil:2013,Glenz:2009zn,Seery:2010kh,Durrer:2009ii, Finelli:2007fr, Marozzi:2011da,Urakawa:2009xaa} and as far as we are aware, this issue is still viewed as unsolved by many \cite{Woodard:2014jba}, see also the recent works \cite{Alinea:2015pza,Alinea:2016qlf,Alinea:2017ncx}.

The approach that implies a significant modification of $\mathcal{P}_\phi$ from renormalization was laid out already in \cite{Parker:1974qw1,Parker:1974qw,Fulling:1974zr} and is generally referred to as adiabatic regularization or adiabatic subtraction. This technique has since been established as one of the most popular approaches for renormalization of quantum fields in curved spaces, for examples see \cite{Bunch:1980vc,Brandenberger:2004kx,Anderson:2005hi,Kaya:2015hka,MolinaParis:2000zz,Habib:1999cs,Anderson:1987yt}. It relies on an adiabatic expansion of modes or more concretely a series in increasing number of derivatives which provides a counter term with the identical divergences to the ones generated by the full quantum correlations, which upon subtraction leads to quantities with the divergences removed. 

Despite the great calculational advantages of adiabatic subtraction it does lack features that in some applications would be desirable. The adiabatic subtraction terms are not in any obvious manner related to redefinitions of (bare) constants as introduced by the action, despite the fact that at the most elementary level that is the essence of renormalizing a quantum theory \cite{Collins:1984xc}. 
The specific physical conditions, in particular having a handle on the renormalization scale and the finite terms in the subtractions that in a strict sense are required for providing physical definitions for the renormalized parameters are similarly only implicitly defined in the procedure. If one wished to make use of a set of counter terms with different finite contributions to the adiabatic prescription and hence a different physical interpretation in terms of matching to observations one would have to go beyond adiabatic subtraction. An approach allowing one to achieve this was recently presented in \cite{Markkanen:2013nwa}. Furthermore, in the works \cite{Cooper:1994ji,Cooper:1996ib,Lampert:1996qw} as summarized in section 14.2.3 of \cite{Calzetta:2008iqa} the adiabatic subtraction terms were converted into redefinitions of couplings in the framework of heavy ion collisions. Similar issues were addressed recently in the context of gravitational waves in \cite{Granese:2017gfb}.
  
In this work by making use of the techniques of \cite{Markkanen:2013nwa} we set out to clarify the non-trivial issues related to the renormalization of the inflationary perturbations and in particular investigate if a modification to the observable spectrum can consistently be kept small even when proper renormalization is implemented.

We will perform our analysis for the simple $m^2\varphi^2$-model given by the matter action
\ee{S_m=-\int d^4x\,\sqrt{-g}\bigg[\f{1}{2}\nabla_\mu\varphi\nabla^\mu\varphi+\f{1}{2}m^2\varphi^2\bigg]\,.\label{eq:act}}
Even though this model is under tension from observations \cite{Ade:2015lrj} it serves as a useful toy model with which to illustrate our calculation in an accessible form. Following \cite{Parker:2007}, we will perform the derivation for $\langle\hat{\phi}^2\rangle$ in the framework of quantum field theory on a curved space time, where the metric is assumed to be a classical background field \cite{Birrell:1982ix,Parker:2009uva}. Although it is strictly speaking not correct to neglect the metric perturbations it can be shown that this approximation will only introduce a small error for the power spectrum.

Our conventions are (+,+,+) \cite{Misner:1974qy} and $c\equiv\hbar\equiv1$.
\section{Adiabatic subtraction}
\label{sec:subtraction}
The adiabatic subtraction prescription is described at length in \cite{Birrell:1982ix,Parker:2009uva} where we refer the reader for more information. In this section we will keep our discussion general by considering a Friedmann--Lema\^{i}tre--Robertson--Walker (FLRW) type metric with
\ee{g_{\mu\nu}dx^\mu dx^\nu=-dt^2+a(t)d\mathbf{x}^2\,,}
with from now on $a(t)\equiv a$, in section \ref{sec:renorm} we will restrict our analysis to de Sitter space exclusively.

Adiabatic subtraction was first discussed in \cite{Parker:1974qw1,Parker:1974qw,Fulling:1974zr} and in practical calculations is probably the most frequently utilized method used for renormalization on a curved background. In this approach a renormalized quantity is defined by subtracting an $A$th order derivative approximation from the full divergent expression. The order of the expansion $A$ depends on the adiabatic order of divergences in the bare quantity and $A$ should only be as high as to include the generated divergences. Thus in adiabatic subtraction all divergences are removed by a single subtraction. Counter terms in the usual sense of redefining constants as introduced by the action are only implicitly defined by the form of the adiabatic subtraction term. Although this brings about a practical advantage it does make the connection to traditional renormalization framework, discussed at length for example in \cite{Collins:1984xc}, less clear. It is however possible also in curved spaces to apply the standard renormalization techniques relying on counter terms derived by redefining bare constants \cite{Markkanen:2013nwa}, which we will discuss in section \ref{sec:renorm}.

In this work we have deliberately chosen to denote this prescription adiabatic subtraction instead of adiabatic regularization despite the fact that it is generally viewed as a regularization method instead of a complete renormalization prescription \cite{Birrell:1982ix,Parker:2009uva}. This is to highlight an important property possessed by it that other regularization methods such as introducing a cut-off or dimensional regularization generally do not: it is not only a method for rendering formally infinite expressions finite, but it by definition contains a prescription for subtraction. When using cut-off or dimensional regularization, in order to complete the renormalization of a quantity one must also define what precisely is to be subtracted from a regularized expression in order to render it physically meaningful. This step one must perform and define {in addition to regularization} and the prescription one chooses defines the finite pieces contained in the subtraction. As an example one can think of the differences in finite terms introduced in the ${\rm MS}$ and $\overline{\rm MS}$ subtraction prescriptions used in particle physics \cite{Collins:1984xc}. This is in fact very important for our purposes: although adiabatic subtraction is usually viewed as a technique with which to regularize divergences it nonetheless implicitly also defines a subtraction prescription including the finite terms in the subtraction. For this reason quantities defined via this technique should be viewed more as renomalized instead of regularized expressions and this distinction will be reflected by our choice of language calling finite quantities defined via adiabatic subtraction renormalized.

We can illustrate the adiabatic prescription with an example: the renormalized variance of the fluctuation of a scalar field we can symbolically define as 
\ea{\label{eq:adsub1}\langle \hat{\phi}^2\rangle_R\equiv\langle \hat{\phi}^2\rangle-(\delta {\phi}^2)^{\rm ad}=\langle \psi\vert \hat{\phi}^2\vert\psi\rangle-\langle 0^{(A)}\vert \hat{\phi}^2\vert 0^{(A)}\rangle\big\vert_{A=2}
}
where $\langle \hat{\phi}^2\rangle$ denotes the bare quantity calculated in some yet undefined state $\vert\psi\rangle$, $(\delta {\phi}^2)^{\rm ad}$ the adiabatic counter term, $\vert0^{(A)}\rangle$ the $A$th order adiabatic vacuum to be defined shortly and we have split the quantized field $\hat{\varphi}$ into a mean field $\langle\hat{\varphi}\rangle$ and a quantum fluctuation
\ee{\hat{\varphi}-\langle\hat{\varphi}\rangle\equiv\hat{\varphi}-{\varphi}\equiv\hat{\phi}\,.\label{eq:fluc}}
The adiabatic subtraction term $(\delta{\phi}^2)^{\rm ad}$ can be derived as follows: first one solves the equation of motion for the fluctuation obtained from (\ref{eq:act})
\ee{\label{eq:eom}\big(-\square+m^2\big){\hat\phi}=0\,,} up to the appropriate adiabatic order.
By using the standard mode decomposition
\ea{\hat{\phi}
=\int d^{n-1}k\,e^{i\mathbf{k}\cdot\mathbf{x}} \big [\hat{a}_\mathbf{k} u_\mathbf{k}+\hat{a}^\dagger_{-\mathbf{k}} u_\mathbf{k}\big],\quad u_\mathbf{k}=\f{h_\mathbf{k}(t)}{\sqrt{(2\pi)^{n-1}a^{n-1}}}\label{eq:ans0}\,,}
the adiabatic solutions come by way of the ansatz
\ee{h^{\rm ad}_\mathbf{k}(t)=\f{1}{\sqrt{2W}}e^{-i\int^{t}Wdt'}\label{eq:ans}\,,} 
where $W$ is expressed as an adiabatic expansion,
\ee{W=c_0+c_1\f{\da}{a}+c_2\f{\da^2}{a^2}+c_3\f{\dda}{a}+\cdots,}
with the $c$'s being functions on $a$, $\mathbf{k}$ and $m$ and where we have normalized our modes such that \ee{[\hat{a}_\mathbf{k},\hat{a}_{\mathbf{k}'}]=[{\hat{a}}^\dagger_\mathbf{k},{\hat{a}}^\dagger_{\mathbf{k}'}]=0,\quad[{\hat{a}}_\mathbf{k},{\hat{a}}^\dagger_{\mathbf{k}'}]=\delta(\mathbf{k}-\mathbf{k}'),} and for completeness analytically continued our dimensions to $n$. 

Explicitly, equation (\ref{eq:eom}) reads
\ee{\bigg[\partial_t\partial_t +(n-1)\f{\da}{a}\partial_t-a^{-2}\partial_i\partial^i+m^2\bigg]\hat{\phi}=0\,,
}
and using the ansatz from (\ref{eq:ans}) gives rise to the equation
\ee{\label{eq:W}W^2=\underbrace{\f{{k}^2}{a^2}+m^2}_{\displaystyle \equiv\omega^2}-\f{\dda}{a}\bigg[\f{1}{2}(n-1)\bigg]-\bigg(\f{\da}{a}\bigg)^2\bigg[\f{1}{4}(n-1)(n-3)\bigg]+\f{3\dot{W}^2}{4W^2}-\f{\ddot{W}}{2W}\,.}
From the above we trivially find the leading adiabatic term to be $(W^2)^{(0)}=\omega^2$ and upon straightforward iteration $(W^2)^{(1)}=0$ and \cite{Markkanen:2013nwa}
\ea{(W^2)^{(2)}&=\frac{\dda}{a}\left[\frac{1}{2}\big(2-n-m^2/\omega^2\big)\right]-\frac{\dot{a}^2}{a^2}\bigg[\f{1}{4}\Big((n-2)^2+4m^2/\omega^2-5\big(m^2/\omega^2\big)^2\Big) \bigg]\label{eq:W2}\,,}
with which the adiabatic counter term for the variance in (\ref{eq:adsub1}) reads
\ea{(\delta \phi^2)^{\rm ad}&=\int \f{d^{n-1}k}{(2\pi a)^{n-1}}|h^{\rm ad}_\mathbf{k}(t)|^2_{A=2}=\int \f{d^{n-1}k}{2(2\pi a)^{n-1}}\bigg[\f{1}{W}\bigg]_{A=2}\nonumber \\&=\int \f{d^{n-1}k}{2(2\pi a)^{n-1}\omega}\bigg\{1-\f{\da^2}{a^2}\bigg[\frac{5 m^4-4 m^2 \omega ^2-(n-2)^2 \omega ^4}{8 \omega ^6}\bigg]+\f{\dda}{a}\bigg[\frac{m^2+(n-2) \omega^2}{4 \omega^4}\bigg]\bigg\}\,.\label{eq:CTa}}
According to the adiabatic prescription the above includes terms with up to two time derivatives only since beyond this order no divergences are generated in four dimensions, which in the above is denoted with $A=2$. 

For inflation in the approximation of strict de Sitter space $a= e^{H t}$ with $H$ almost a constant the equation of motion (\ref{eq:eom}) can be solved exactly. Making the standard and well-motivated choice of the Bunch-Davies (BD) vacuum as the boundary condition for the mode \cite{Chernikov:1968zm,BD} and focussing on the four-dimensional case the solution is given as
\ee{h_{\mathbf{k}}(t)=\sqrt{\f{\pi}{4H}}H^{(1)}_{{\nu}}\big(k/(aH)\big)\,; \qquad\nu^2=\f{9}{4}-\f{m^2}{H^2}\label{eq:sol}\,.}
Using the definitions (\ref{eq:Rp0}) and (\ref{eq:adsub1}) for the BD mode as well as the adiabatic counter term (\ref{eq:CTa}) one may easily find an expression for the adiabatically renormalized power spectrum for the field to be
\ea{\label{eq:Rp}\mathcal{P}_\phi-\delta \mathcal{P}^{\rm ad}_\phi&\equiv(\mathcal{P}_\phi)_R=\f{(k/a)^3}{2\pi^2}\Big(|h_\mathbf{k}(t)|^2-|h^{\rm ad}_\mathbf{k}(t)|_{A=2}^2\Big)\nonumber \\&=\bigg(\f{k}{a H}\bigg)^3\f{H^2}{8\pi}\bigg\{\big|H^{(1)}_{{\nu}}\big(k/(aH)\big)\big|^2-\bigg[\frac{2 H}{\pi  \omega}-\frac{H^3 \left(5 m^4-6 m^2 \omega ^2-8 \omega^4\right)}{4 \pi  \omega^7}\bigg]\bigg\}\,,}
which for long wavelengths, ${k}/({aH})\longrightarrow0$, and small masses, $\nu\longrightarrow3/2$,
gives \ee{(\mathcal{P}_\phi)_R\longrightarrow \bigg(\f{H}{2\pi}\bigg)^2\bigg\{1-\bigg(\f{k}{a H}\bigg)^3\bigg[\frac{H}{\omega}-\frac{H^3 \left(5 m^4-6 m^2 \omega^2-8 \omega^4\right)}{8 \omega ^7}\bigg]\bigg\}\,.}
As discovered in \cite{Parker:2007} for small masses (\ref{eq:Rp}) deviates from the unrenormalized expression by several orders of magnitude. For example for $m^2/H^2\sim 10^{-2}$ one gets
\ee{\f{(\mathcal{P}_\phi)_R}{\mathcal{P}_\phi}\sim 10^{-4} \,,\label{eq:mods}}
for scales exiting the horizon, $k/(aH)\sim1$. This obviously implies a significant modification to the standard predictions. 

An important point to keep in mind however is that even if (\ref{eq:mods}) is true in the adiabatic prescription, it does not imply that a large correction from renormalization is always unavoidable:
it is well-known from the context of particle theory \cite{Peskin:1995ev} that the counter terms rendering a theory finite have universal divergencies but in general may have different finite contributions depending on the physical boundary conditions one imposes and as discussed in the beginning of this section, adiabatic subtraction is simply one prescription for defining the finite parts of the counter terms. In principle, nothing prevents the existence of other renormalization prescriptions where the long wavelength portion of the spectrum is only mildly modified. Also, the renormalization scale imposed in the adiabatic subtraction terms cannot be easily determined thus rendering the physical interpretation of the subtraction non-trivial.

The above assertion that there must exist other prescriptions beyond adiabatic subtraction is also implied by the axiomatic approach laid out in \cite{Wald:1977up,Wald:1978pj,Wald:1978ce} by Wald: under a few physically motivated criteria the renormalized energy-momentum tensor is unique only up to a covariantly conserved local tensor. For more discussion of this axiomatic approach, see for example chapter 6.6 of \cite{Birrell:1982ix}.

Furthermore, although for a non-interacting theory the terms introduced by adiabatic subtraction can be reduced to a redefinition of the bare constants of the action, which in principle is ultimately what one does when renormalizing, in the interacting case this no longer applies \cite{Markkanen:2013nwa}. Hence, when interactions are included adiabatic subtraction cannot, in a strict sense, be considered a consistent renormalization prescription. Let us then turn to the technique introduced in \cite{Markkanen:2013nwa} where these issues are not present and where one may easily choose arbitrary finite pieces for the counter terms.


\section{Counter terms in curved space}
\label{sec:renorm}
In this section we will throughout make use of the de Sitter approximation with $a=e^{H t}$ and $\dot{H}=0$.

One of the main difficulties in performing consistent renormalization on a curved background is respecting coordinate invariance: the renormalized energy-momentum tensor is required to be covariantly conserved which directly implies the same for the respective counter term
\ee{\nabla^\mu\langle\hat{T}_{\mu\nu}\rangle_R\equiv\nabla^\mu\big(\langle\hat{T}_{\mu\nu}\rangle-\delta T_{\mu\nu}\big)=0\qquad\Leftrightarrow\qquad \nabla^\mu\delta T_{\mu\nu}=0\,.\label{eq:covCons1}}
Trying to \textit{a priori} devise a counter term $\delta T_{\mu\nu}$ that removes all generated divergences while at the same time respects the above requirement and does not subtract physically relevant contributions is generally quite non-trivial as (\ref{eq:covCons1}) imposes constraints on the form of the subtractions as well as the regularization used \cite{Birrell:1982ix,Parker:2009uva}. In adiabatic subtraction no issues arise since the adiabatic modes (\ref{eq:ans}) satisfy the equation of motion up to the required truncation which implies that covariant conservation must be respected. Furthermore, since the subtraction term is given implicitly by an integral as in (\ref{eq:CTa}) one may perform the subtraction procedure by simply combining two formally divergent expressions under the same integral and the issues related to properly regulating the integrals do not arise. This is true also when interactions are present \cite{MolinaParis:2000zz}.
  
However, as discussed in the previous section adiabatic subtraction has two features that in some applications render it less ideal: first, as demonstrated in \cite{Markkanen:2013nwa} in the interacting case it does not correspond to redefining the parameters of the theory into counter terms and finite physical contributions. Second, it does not posses a handle with which to control the finite pieces of the subtraction. These issues are evaded when implementing the technique of \cite{Markkanen:2013nwa}.

Any renormalization prescription in a strict sense should in the end be traced back to a redefinition of constants in the original action i.e. all necessary subtractions should be obtainable from counter terms resulting from the appropriate redefinition of the bare constants, $c_i$, into finite physical pieces, $c_{i,R}$, and divergent counter terms, $\delta c_i$, as
\ee{c_i\longrightarrow c_{i,R}+\delta c_i\,,\label{eq:cts}}
which once determined for example from the renormalization of the energy-momentum tensor must remove all divergences that appear in physical quantities.
Importantly, when renormalization is performed by introducing counter terms via (\ref{eq:cts}) the covariant conservation requirement (\ref{eq:covCons1}) is automatically satisfied given that general coordinate invariance is not broken by an improper choice of regularization, such as a cut-off \cite{Maggiore:2010wr2,Akhmedov:2002ts}. Specifically, analytic continuation to $n$ dimensions allows one to write closed expressions for the counter terms resulting from (\ref{eq:cts}) while respecting (\ref{eq:covCons1}), as for example implemented in \cite{Markkanen:2013nwa}. In this work we will also highlight this feature with an example calculation (see equations (\ref{eq:l1} -- \ref{eq:l2})). 

In a first principle approach to renormalization in curved space relying on (\ref{eq:cts}) it is well-known that in order to have all necessary counter terms one must also introduce the following gravitational action\footnote{Often the cosmological constant is defined to have mass dimension two, which is obtained from our definition via $\Lambda\rightarrow M_{\rm P}^2\Lambda$} \cite{Birrell:1982ix,Parker:2009uva}
\ee{\label{eq:actg}S_g= \f{1}{2}\int d^4x\sqrt{-g}~\bigg[-2\Lambda +{M_{\rm P}^2} R+\beta R^2+\epsilon_{1}R_{\alpha\beta}R^{\alpha\beta}+ \epsilon_{2}R_{\alpha\beta\gamma\delta}R^{\alpha\beta\gamma\delta}\bigg]
\,,}
in addition to the matter piece in (\ref{eq:act}) i.e. \ee{S=S_m+S_g\,.}
The first two terms in (\ref{eq:actg}) are the familiar cosmological constant and Einstein-Hilbert term with the reduced Planck mass, $M_{\rm P}^2\equiv (8\pi G)^{-1}$, where however the parameters are not yet physical as the redefinitions (\ref{eq:cts}) have not yet been made.

In the semi-classical approach with quantum matter and a classical metric $S_g$ provides counter terms on the matter or energy-momentum side of the Einstein equation, despite the fact that the tree-level contributions from $S_g$ only appear in on the geometric side of the semi-classical Einstein equation
\ea{\label{eq:renomE0}\f{2}{\sqrt{-g}}\f{\delta S_g}{\delta g^{\mu\nu}}&=\langle\hat{T}_{\mu\nu}\rangle_R\,;\\\langle\hat{T}_{\mu\nu}\rangle_R&\equiv\langle\hat{T}_{\mu\nu}\rangle-\delta T_{\mu\nu}=-\bigg\langle\f{2}{\sqrt{-g}}\f{\delta S_m}{\delta g^{\mu\nu}}\bigg\rangle-\f{2}{\sqrt{-g}}\f{\delta S_g}{\delta g^{\mu\nu}}\bigg\vert_{c_i\rightarrow\delta c_i}\,,}
where for a non-interacting theory the only counter terms required come from $S_g$. In an interacting theory counter terms from $S_m$ also need to be included \cite{Markkanen:2013nwa}.  

The energy-momentum tensor can be conveniently split into a sum of the classical, quantum and counter term pieces, $\langle\hat{T}_{\mu\nu}\rangle_R\equiv T_{\mu\nu}^C+\langle\hat{T}^Q_{\mu\nu}\rangle-\delta T_{\mu\nu}$, which for our theory with (\ref{eq:act}) and (\ref{eq:actg}) are
\ea{\langle\hat{T}_{\mu\nu}\rangle_R&=-\f{g_{\mu\nu}}{2}\Big[\partial_\rho\varphi\partial^\rho\varphi+m^2_R\varphi^2\Big]+\partial_\mu\varphi\partial_\nu\varphi\\&+\label{eq:renomE1}\bigg\langle-\f{g_{\mu\nu}}{2}\Big[\partial_\rho\hat{\phi}\partial^\rho\hat{\phi} +m^2_R\hat{\phi}^2\Big]+\partial_\mu\hat{\phi}\partial_\nu\hat{\phi}\bigg\rangle\\ &-\Big[g_{\mu\nu}\delta\Lambda+\delta M_{\rm P}^2G_{\mu\nu} +\delta\beta~^{(1)}H_{\mu\nu}+\delta\epsilon_1~^{(2)}H_{\mu\nu}+\delta\epsilon_2H_{\mu\nu}\Big]
\,,\label{eq:renomE2}}
where in $n$-dimensional de Sitter space we have \cite{Markkanen:2013nwa}
\begin{align}
G_{\mu\nu}&\equiv\f{1}{\sqrt{-g}}\f{\delta}{\delta g^{\mu\nu}}\int d^nx\sqrt{-g}~R^{\phantom{2}}=g_{\mu\nu}\frac{1}{2}(2-n) (n-1) H^2 \,,\\
~^{(1)}H_{\mu\nu}&\equiv\f{1}{\sqrt{-g}}\f{\delta}{\delta g^{\mu\nu}}\int d^nx\sqrt{-g}~R^2= 
g_{\mu\nu}\frac{n}{2} (4-n) (n-1)^2  H^4\,,\label{eq:h4}
\\
~^{(2)}H_{\mu\nu}&\equiv\f{1}{\sqrt{-g}}\f{\delta}{\delta g^{\mu\nu}}\int d^nx\sqrt{-g}~R_{\mu\nu}R^{\mu\nu}= g_{\mu\nu}\frac{1}{2} (4-n) (n-1)^2  H^4\,,
\\ 
H_{\mu\nu}&\equiv\f{1}{\sqrt{-g}}\f{\delta}{\delta g^{\mu\nu}}\int d^nx\sqrt{-g}~R_{\alpha\beta\gamma\delta}R^{\alpha\beta\gamma\delta}=  g_{\mu\nu} (4-n) (n-1)  H^4 \label{eq:h42}\,.
\end{align}
As one may see from (\ref{eq:h4} - \ref{eq:h42}), in (\ref{eq:renomE2}) in de Sitter space one only needs to include one of the $\mathcal{O}(R^2)$ tensors in order to cancel all divergences $\propto H^4$ and from now on we will neglect the counter terms from the contributions $R_{\alpha\beta}R^{\alpha\beta}$ and $R_{\alpha\beta\gamma\delta}R^{\alpha\beta\gamma\delta}$ by simply setting $\delta\epsilon_1=\delta\epsilon_2=0$\footnote{In a general FLRW background this cannot be done since$~^{(1)}H_{\mu\nu}$,$~^{(2)}H_{\mu\nu}$ and $H_{\mu\nu}$ all have distinct expressions in terms of $\da/a$, $\dda/a$, $\ddda/a$ and $a^{(4)}/a$ with dimension $H^4$ \cite{Markkanen:2013nwa}.}.

Before discussing the specifics of our method, we comment on an important and non-trivial feature of renormalization on a curved background that we already briefly mentioned at the beginning of this section. In order to verify that the divergences in the energy-momentum tensor do in fact organize themselves as a linear combination of local and conserved curvature tensors as in (\ref{eq:renomE2}) a proper regularization method must be introduced. On a curved background some regulators such as a cut-off are problematic as they lead to explicit violation of covariance and hence render any renormalization approach based on redefinition of constants more involved. Specifically, when using a cut-off some non-covariant divergences must simply by removed by hand \cite{Maggiore:2010wr2,Akhmedov:2002ts}. Fortunately, in dimensional regularization this problem does not surface which is the reason why throughout we have maintained $n$ dimensions in our expressions. In dimensional regularization covariance is often maintained in a very non-trivial manner related to the special properties of the analytically continued integrals.

One may clarify the above discussion by calculating the vacuum energy and pressure densities in flat space, which come easily with the mode (\ref{eq:ans0}) inserted into (\ref{eq:renomE1}) with simply $W^2 = {k}^2+m^2_R$
\ea{\langle\hat{T}^Q_{00}\rangle_{H=0}&=\f{1}{2}\int \frac{d^{n-1 }{k}}{(2\pi)^{n-1}}{\sqrt{{k}^2+m^2_R}}=\f{1}{2(4\pi)^{\f{n-1}{2}}}\f{\Gamma[-\f{n}{2}]}{\Gamma[-\f{1}{2}]}m^n_R\nonumber \\&=\frac{m^4_R}{32 \pi ^2 (n-4)}+\frac{m^4_R}{128 \pi
		^2} \left(2 \log \left(\frac{m^2_R}{4 \pi }\right)+2 \gamma -3\right)+\mathcal{O}(n-4)\,,\label{eq:l1}} for the energy density and for the pressure
\ea{\langle\hat{T}^Q_{ii}\rangle_{H=0}&=\f{1}{2}\int \frac{d^{n-1 }{k}}{(2\pi)^{n-1}}\frac{{k}^2}{(n-1)\sqrt{{k}^2+m^2_R}}=\f{1/2}{2(4\pi)^{\f{n-1}{2}}}\f{\Gamma[-\f{n}{2}]}{\Gamma[\f{1}{2}]}m_R^{n}\nonumber \\&=-\frac{m^4_R}{32 \pi ^2 (n-4)}-\frac{m^4_R}{128 \pi
		^2} \left(2 \log \left(\frac{m^2_R}{4 \pi }\right)+2 \gamma -3\right)+\mathcal{O}(n-4)\,,\label{eq:l2}}
where the final expressions in (\ref{eq:l1} -- \ref{eq:l2}) can be obtained with the standard formulae of dimensional regularization \cite{Peskin:1995ev}. Quite obviously (\ref{eq:l1} -- \ref{eq:l2}) satisfy \ee{\langle\hat{T}^Q_{\mu\nu}\rangle_{H=0}\propto g_{\mu\nu}\label{eq:porp}} as required by (\ref{eq:renomE1} -- \ref{eq:renomE2}). The fact that all divergences generated for the energy-momentum tensor when dimensionally regularized organize themselves in the form (\ref{eq:renomE2}) was verified in \cite{Markkanen:2013nwa}, in a general FLRW background\footnote{See in particular equation (2.14), Appendix B and the discussion in section 5.2 of \cite{Markkanen:2013nwa}.}. What is also apparent is that for a cut-off (\ref{eq:porp}) fails since, as we already mentioned, a cut-off violates covariance. This also implies that in a precise sense the integrals are valid as $n$-dimensional expressions and only if the integrand for some expression is convergent in four dimensions can one set $n=4$.

In the adiabatic prescription the quantum piece of the energy-momentum tensor can be renormalized as was done in (\ref{eq:adsub1}) for the variance
\ea{\label{eq:adsub2}
\langle \hat{T}^Q_{\mu\nu}\rangle_R&= \langle\hat{T}^Q_{\mu\nu}\rangle-\delta T^{\rm ad}_{\mu\nu}\equiv\langle\psi| \hat{T}^Q_{\mu\nu}|\psi\rangle-\langle 0^{(A)}\vert \hat{T}^Q_{\mu\nu}\vert 0^{(A)}\rangle\big\vert_{A=4}\,,
}
where in four dimensions the adiabatic expansion in general has to be performed up to fourth order in derivatives in order to remove all divergences. As we discussed in the previous section adiabatic subtraction is just one choice for the finite parts of the counter terms $\delta\Lambda$, $\delta M_{\rm P}^2$ and $\delta \beta$. However, since the divergencies are universal adiabatic subtraction can be used to determine the divergent parts of the counter terms. This leads to the following modified approach with which one may introduce \textit{arbitrary} finite contributions in the counter terms: if we match the adiabatic counter term in (\ref{eq:adsub2}) with the general form in given in (\ref{eq:renomE2}) as
\ee{\langle 0^{(A)}\vert \hat{T}^Q_{\mu\nu}\vert 0^{(A)}\rangle\big\vert_{A=4}\equiv \delta T^{\rm ad}_{\mu\nu}=g_{\mu\nu}\delta\Lambda^{\rm ad}+\delta (M_{\rm P}^{2})^{\rm ad}G_{\mu\nu} +\delta\beta^{\rm ad}~^{(1)}H_{\mu\nu}\label{eq:match}\,,}
we may extract the specific choice made implicitly for the counter terms in the adiabatic prescription. As is also apparent in the results of the previous section, adiabatic subtraction results in counter terms that are given in terms of momentum space integrals and once we have explicit expressions for $\delta\Lambda^{\rm ad}$, $\delta (M_{\rm P}^{2})^{\rm ad}$ and $\delta\beta^{\rm ad}$ we may straightforwardly find other sets of counter terms with identical divergences but with differing finite parts by changing the parts that remain finite when $k/a\rightarrow\infty$. 
We also note that it has been shown that adiabatic subtraction, including the generated finite pieces, does admit an expression of the form (\ref{eq:match}) \cite{Markkanen:2013nwa}.

The expression for energy-momentum can be read from (\ref{eq:renomE1}) with which we can solve the expressions for the energy and pressure density adiabatic counter terms as 
\ee{ \delta\hat{T}^{\rm ad}_{00}=\f{1}{2}\int d^{n-1} {k}\bigg[\vert\dot{u}^{\rm ad}_\mathbf{k}\vert^2+\bigg( \f{\mathbf{k}^2}{a^2}+m_R^2\bigg)\vert u^{\rm ad}_\mathbf{k}\vert^2\bigg]\label{eq:T00}}
and
\ee{
a^{-2}\delta\hat{T}^{\rm ad}_{ii}=\f{1}{2}\int d^{n-1} k\bigg[\vert\dot{u}^{\rm ad}_\mathbf{k}\vert^2-\bigg(\f{3-n}{1-n}\f{\mathbf{k}^2}{a^2}+m_R^2\bigg)\vert u^{\rm ad}_\mathbf{k}\vert^2\bigg]\label{eq:Tii}\,,}
again for completeness in $n$ dimensions. Making use of  (\ref{eq:ans}) and (\ref{eq:match}) along with (\ref{eq:T00}) and (\ref{eq:Tii}) gives
\ea{\delta\hat{T}^{\rm ad}_{00}-a^{-2}\delta\hat{T}^{\rm ad}_{ii}&=-2\delta\Lambda^{\rm ad}+\delta (M_{\rm P}^{2})^{\rm ad}\big(G_{00}-a^{-2}G_{ii}\big) +\delta\beta^{\rm ad}\big(^{(1)}H_{00}-a^{-2}~^{(1)}H_{ii}\big)\label{eq:je} \\ &=\int \f{d^{n-1}k}{2(2\pi a)^{n-1}}\bigg[\f{1}{W}\bigg(\f{2-n}{1-n}\f{\mathbf{k}^2}{a^2}+m_R^2\bigg) \bigg]_{A=4}\,,}
which now requires one to find a solution for $W$ to fourth adiabatic order, which comes via a straightforward but tedious iteration of equation (\ref{eq:W}). With the help of the appendixes of \cite{Markkanen:2013nwa} and \cite{Markkanen:2016aes}, after some work, we can solve the counter terms as defined in the adiabatic prescription to be
\ea{\label{eq:della}\delta\Lambda^{\rm ad}&\equiv\int\f{d^{n-1}k}{(2\pi a)^{n-1}}\delta\Lambda^{\rm ad}_{\mathbf{k}}=\int\f{d^{n-1}k}{(2\pi a)^{n-1}}\frac{m_R^2+(n-2) \omega ^2}{4 (1-n) \omega }\,,\\\label{eq:dellab}\delta (M_{\rm P}^{2})^{\rm ad}&\equiv\int\f{d^{n-1}k}{(2\pi a)^{n-1}}\delta(M_{\rm P}^{2})^{\rm ad}_{\mathbf{k}} =\int \f{d^{n-1}k}{(2\pi a)^{n-1}}\frac{\left(m_R^2+(n-2) \omega ^2\right)}{16 (n-2) (n-1)^2 \omega ^7} \bigg\{-5 m_R^4\\ \nonumber&+6 m_R^2 \omega ^2+(n-2) n \omega ^4\bigg\}\,,}
and
\ea{\delta\beta^{\rm ad}&\equiv\int\f{d^{n-1}k}{(2\pi a)^{n-1}}\delta\beta^{\rm ad}_{\mathbf{k}}=\int\f{d^{n-1}k}{(2\pi a)^{n-1}}\frac{m_R^2+(n-2) \omega^2}{256 (n-4) (n-1)^3 n \omega^{13}}\bigg\{1155 m_R^8-2772 m_R^6 \omega ^2\nonumber \\&-70 m_R^4 ((n-2) n-30) \omega ^4+20 m_R^2 (5 (n-2) n-24) \omega ^6+3 (n-4) (n^2-4) n\omega ^8\bigg\}\,.\label{eq:delbe}}
We note that all dependence on the scale factor in (\ref{eq:della} - \ref{eq:delbe}) is of the form $k/a$ as it only comes from the $a$-dependence of the integration measure and $\omega^2=(k/a)^2+m^2$. Hence, by redefining the integration variable
\ee{\f{{k}}{a}\equiv {q}\qquad \Rightarrow\qquad \f{d^{n-1}k}{(2\pi a)^{n-1}}\equiv\f{d^{n-1}q}{(2\pi )^{n-1}}\label{eq:intv}\,,}
(\ref{eq:della} - \ref{eq:delbe}) can be seen to be strictly constant. By again using standard formulae of dimensional regularization \cite{Peskin:1995ev} one may check that the divergencies of $\delta\Lambda^{\rm ad}$ and $\delta (M_{\rm P}^{2})^{\rm ad}$ match the well-known results, for example as calculated by other means in section 6.2 of \cite{Birrell:1982ix}\footnote{\ee{\delta\Lambda^{\rm ad}=\frac{m_R^4}{32 \pi ^2 (4-n)}+\cdots\,;\qquad\delta (M_{\rm P}^{2})^{\rm ad}=\frac{m_R^2}{48 \pi ^2 (4-n)}+\cdots} As we have made use of simplifications arising in strict de Sitter space a direct correspondence with the results of \cite{Birrell:1982ix} and $\delta\beta^{\rm ad}$ is less obvious.}.

In (\ref{eq:della} - \ref{eq:delbe}) we have now successfully extracted the constant counter terms that are implicitly defined by adiabatic subtraction. We emphasize once more that only the divergent parts in (\ref{eq:della} - \ref{eq:delbe}) are universal and finite contributions, or terms in the integrands that vanish faster than $\sim q^{-4}$ at the high ultraviolet when $n=4$, should be determined according to the dedicated renormalization prescription of one's choosing. Note that since we are interested in the four-dimensional results, the last term in (\ref{eq:je}) only generates a finite contribution, since from (\ref{eq:h4}) one has$~^{(1)}H_{\mu\nu}\sim (4-n)H^4$ and by explicitly performing the integral in (\ref{eq:delbe}) one can show that the leading term in $\delta\beta^{\rm ad}$ when approaching $n\rightarrow4$ scales as $(4-n)^{-1}$.

With (\ref{eq:della} - \ref{eq:delbe}) we can now address the main focus of this article, which is finding renormalization prescriptions where the long wavelength limit of the renormalized power spectrum $(\mathcal{P}_\phi)_R$ is much less affected than in the adiabatic approach in (\ref{eq:Rp}).

\section{Renormalization of the spectrum of perturbations}
\label{sec:modp}
In order to derive a renormalized expression for the spectrum of perturbations from inflation as defined in (\ref{eq:R}) we need to find a physical quantity that is a function of $\langle\hat{\phi}^2\rangle$. After all, a physical expression involving $\langle\hat{\phi}^2\rangle$ must also include the appropriate counter terms providing a way of defining $\delta\mathcal{P}_\phi$. We need to look no further than the Einstein equations, since by making use of expressions (\ref{eq:T00}) and (\ref{eq:Tii}), but for the full mode $u_\mathbf{k}^{\rm ad}\rightarrow u_\mathbf{k}^{\phantom{\rm ad}}$ and with the definition for the spectrum (\ref{eq:PS}) we get
\ee{\langle \hat{T}^Q_{00}\rangle-a^{-2}\langle \hat{T}^Q_{ii}\rangle=\int_0^\infty \f{k^{n-2}dk}{(2\pi a)^{n-1}}\bigg(\f{2-n}{1-n}\f{{k}^2}{a^2}+m_R^2\bigg)\f{\mathcal{P_\phi}}{(k/a)^{n-1}}\,.\label{eq:toi}} 
Equation (\ref{eq:toi}) after the appropriate renormalization we can then use as the \textit{definition} of the renormalized power spectrum 
\ea{\langle\hat{T}^Q_{00}\rangle_R-a^{-2}\langle \hat{T}^Q_{ii}\rangle_R&=\langle \hat{T}^Q_{00}\rangle-a^{-2}\langle \hat{T}^Q_{ii}\rangle\nonumber \\&-\big[-2\delta\Lambda+\delta M_{\rm P}^{2}\big(G_{00}-a^{-2}G_{ii}\big) +\delta\beta\big(^{(1)}H_{00}-a^{-2}~^{(1)}H_{ii}\big)\big]\nonumber \\ &\equiv \int_0^\infty \f{k^{n-2}dk}{(2\pi a)^{n-1}}\bigg(\f{2-n}{1-n}\f{{k}^2}{a^2}+m_R^2\bigg)\f{\mathcal{P_\phi}-\delta \mathcal{P_\phi}}{(k/a)^{n-1}}\,.\label{eq:physP}}
As an example we can use the momentum space expressions for the adiabatic counter terms for $\delta\Lambda^{\rm ad}$ and $\delta(M_{\rm P}^2)^{\rm ad}$, furthermore setting $\delta\beta=0$ and finally solve $(\mathcal{P}_\phi)_R$ from (\ref{eq:physP}) in $n=4$ to be
\ea{\mathcal{P_\phi}-\delta \mathcal{P}_\phi^{\rm ad}\equiv(\mathcal{P}_\phi)_R&=\bigg(\f{k}{a H}\bigg)^3\f{H^2}{8\pi}\bigg\{\big|H^{(1)}_{{\nu}}\big(k/(aH)\big)\big|^2 \nonumber \\&-H(8\pi)^2\bigg(\f{2}{3}\f{{k}^2}{a^2}+m_R^2\bigg)^{-1}\Big[-\delta\Lambda^{\rm ad}_\mathbf{k}+\delta (M_{\rm P}^{2})^{\rm ad}_\mathbf{k} 3H^2\Big]\bigg\}\,,\label{eq:adpi}}
which after plugging in the explicit results for $\delta\Lambda^{\rm ad}_\mathbf{k}$ and $\delta(M_{\rm P}^2)^{\rm ad}_\mathbf{k}$ from (\ref{eq:della}) and (\ref{eq:dellab}) precisely coincides with the result of the derivation of section \ref{sec:subtraction} given in equation (\ref{eq:Rp}). The reason why we needed to choose $\delta\beta=0$ in order to find agreement with adiabatic subtraction is that the variance generates divergences only to second adiabatic order and in the adiabatic prescription the subtraction term is truncated to neglect the orders beyond the divergences. However, the energy-momentum tensor generically has divergences up to fourth order and since we used the energy-momentum tensor to define the physical $(\mathcal{P}_\phi)_R$ in our approach a dependence on counter terms of $\mathcal{O}(H^4)$ is introduced, although $\delta\beta$ is strictly not required to cancel any divergences in $\mathcal{P}_\phi$.

Before proceeding let us first discuss a few key elements of our approach and the definition for the renormalization of the power spectrum $(\mathcal{P}_\phi)_R$ we introduced in (\ref{eq:physP}). Comparing the adiabatic prescription of section \ref{sec:subtraction} and the result (\ref{eq:adpi}) it seems we have only been able to derive the identical results in a different form, but in a seemingly more cumbersome manner. The significance of our approach however lies in the fact that throughout we maintain a connection with counter terms in the traditional sense as the renormalized spectrum depends explicitly on coupling constants introduced by the tree-level action. Namely, we succesfully mapped the renormalization of the power spectrum into redefinitions of the cosmological constant and the Planck mass, with the simplifying choice of $\delta\beta=0$. By deciphering the counter term content implicitly defined by adiabatic subtraction in (\ref{eq:della} - \ref{eq:delbe}) then allows us to make a leap forward: any set of counter terms that coincides with (\ref{eq:della} - \ref{eq:delbe}) in terms of the divergences represents in principle a valid renormalization prescription. As the explicit expressions (\ref{eq:della} - \ref{eq:delbe}) are given as momentum space integrals modifying the finite terms is straightforward, as we will now show.

Suppose that instead of the adiabatic counter terms (\ref{eq:della} - \ref{eq:delbe}) we made use of the following choices
\ea{\label{eq:della0}\delta\underline{\Lambda}\equiv\int\f{d^{n-1}k}{(2\pi a)^{n-1}}e^{-(aM/k)^5}\delta\Lambda^{\rm ad}_{\mathbf{k}}\,,\qquad\delta \underline{M}_{\rm P}^{2}\equiv\int\f{d^{n-1}k}{(2\pi a)^{n-1}}e^{-(aM/k)^5}\delta(M_{\rm P}^{2})^{\rm ad}_{\mathbf{k}}\,,\qquad \delta\underline{\beta}=0\,,}
where $M$ is some large mass scale. For $n=4$ the leading infinities in (\ref{eq:della} - \ref{eq:delbe}) come from terms $\sim\int^\infty dq\, q^3$ where again the time dependence has disappeared by way of (\ref{eq:intv}) and hence the above counter terms generate precisely the same divergences as (\ref{eq:della} - \ref{eq:delbe}). However, they become quickly suppressed for $q<M$ and the observable modes in $\mathcal{P}_\phi$ are unaltered in this renormalization prescription. Since (\ref{eq:della0}) correspond to constant redefinitions of the cosmological constant and the Planck mass and furthermore we have not introduced a cut-off but give the counter terms as formally divergent integrals covariant conservation of the renormalized energy-momentum tensor (\ref{eq:covCons1}) must hold. Furthermore, one may calculate the difference in renormalization prescriptions for the energy-momentum tensor between the adiabatic and modified approach by making use of (\ref{eq:match}), (\ref{eq:della} - \ref{eq:delbe}) and (\ref{eq:della0}) and subtracting the adiabatic result from the modified one, which results in a finite contribution formed as a linear combination of only local and conserved curvature tensors, $g_{\mu\nu},\, G_{\mu\nu}$ and ${\,}^{(1)}H_{\mu\nu}$. This is in accord with the ambiguity allowed by the axiomatic approach, as discussed at the end of section \ref{sec:subtraction}. Finally, in regards to our discussion after equation (\ref{eq:porp}) since the prescription (\ref{eq:della0}) was defined by modifying the adiabatic expressions, there is no danger of generating new divergences that require the $n$-dimensional forms of the integrals and one may consistently set $n=4$.  

Making the choices in (\ref{eq:della0}) will via (\ref{eq:physP}) give rise to a renormalized power spectrum which explicitly reads
\ea{\mathcal{P}_\phi-\delta \underline{\mathcal{P}}_\phi&\equiv (\mathcal{P}_\phi)_R=\bigg(\f{k}{a H}\bigg)^3\f{H^2}{8\pi}\bigg\{\big|H^{(1)}_{{\nu}}\big(k/(aH)\big)\big|^2\nonumber \\&-e^{-(aM/k)^5}\bigg[\frac{2 H}{\pi  \omega}-\frac{H^3 \left(5 m_R^4-6 m_R^2 \omega ^2-8 \omega^4\right)}{4 \pi  \omega^7}\bigg]\bigg\}\,;\qquad \omega^2\equiv \f{k^2}{a^2}+m_R^2\,,\label{eq:Rpr}}
and which trivially coincides with the bare results for the observable modes, in particular
\ee{(\mathcal{P}_\phi)_R\longrightarrow \bigg(\f{H}{2\pi}\bigg)^2\,;\quad \text{for}\quad \f{k}{aH}\longrightarrow0~~\text{and}~~\nu\longrightarrow3/2\,,}
without compromising the successful removal of the ultraviolet divergences or coordinate invariance of the theory.

As a final comment, we stress that the counter terms in (\ref{eq:della0}) serve merely to illustrate that there exist consistent renormalization prescriptions where the observable modes in the spectrum of inflationary perturbations are not affected, contrary to the claim of \cite{Parker:2007}. 
A different question all together is providing a set of renormalization conditions based on observations corresponding to clear physical definitions for the cosmological constant and the Planck mass. However, this question can also be approached with the techniques presented here and in \cite{Markkanen:2013nwa}.

\section{Discussion}
In works investigating Early Universe phenomenology, in particular the physics and implications of cosmological inflation, renormalization in any form is rarely included in the discussion which stems perhaps from the implicit assumption that observable predictions from the bare theory are not significantly altered by it. Although renormalizing a quantum theory on a curved background is known to be a non-trivial endeavour, for any theory containing divergent correlators some consistent procedure for removing the infinities must in principle be applied in order to obtain robust results.

If the subtractions involved in devising a well-defined finite theory in curved space unavoidably gave rise to a significant change over the bare results as suggested in \cite{Parker:2007} it would force the community to completely overhaul the standard framework with which observable predictions from inflation are derived. This compels a closer look on the process of renormalization in curved space and specifically its effect on the amplitude of perturbations from inflation which was the primary source of the claim in \cite{Parker:2007} and the debate that followed.

The specific prescription which when implemented on the renormalization of inflationary perturbations significantly modifies the bare results is adiabatic subtraction or regularization, a technique which does not immediately connect to the standard approaches used in the context of particle physics: it does not rely on explicitly redefining the coupling constants of the action to produce the necessary counter terms with which the infinities are removed. It also does not seem to require any information about the specific conditions one needs to choose for defining the physical couplings. This leaves the choices made for finite pieces in the subtraction, which in any subtraction prescription are unavoidable, completely implicit. A technique that can overcome these issues was laid out recently in \cite{Markkanen:2013nwa}.

In this work by making use of the approach of \cite{Markkanen:2013nwa} we reduced the procedure of adiabatic subtraction into a redefinition of constants allowing one to easily connect it to the more traditional approaches of renormalization. From the renormalized energy-momentum we were able to define the renormalized power spectrum in a way where the subtraction terms were directly mapped to redefinitions of the bare cosmological constant and Planck mass. Disentangling the implicit definitions for the counter terms made in the adiabatic prescription allowed us to find other prescriptions with relative ease, which differ from one another by only local and conserved tensors and hence fall within the ambiguity allowed by the axiomatic approach of \cite{Wald:1977up,Wald:1978pj,Wald:1978ce}. Finally, we showed how one may renormalize the power spectrum such that it is observationally indistinguishable from the prediction of the bare theory.

Our main result is likely to be viewed favourably by the majority of the community working on Early Universe physics as it implies that the observable spectrum of perturbations is not necessarily modified by renormalization, which is the mainstream assumption. However, making any stronger claims at this stage would be somewhat premature: we have showed that prescriptions exist where this is true, but it is a different question to define a well-motivated physical condition for renormalization where the same continues to hold. It is also non-trivial to generalize our approach to the case of an interacting theory as well as to a case where the background is not approximated to be strictly de Sitter, which also would require the inclusion of the metric fluctuations into the calculation. 

To conclude,  we mention a potentially fruitful future line of research implied by our results. The much used stochastic approach \cite{Starobinsky:1994bd,Starobinsky:1986fx} for determining the dynamics of scalar fields in quasi-de Sitter space includes no explicit discussion of renormalization. Hence it would be interesting to explore the consequences from various different renormalization prescriptions on the stochastic approach and their physical interpretations. A specific question that seems worth investigating is whether the important features such as the attractor nature \cite{Grain:2017dqa} or equilibration \cite{Hardwick:2017fjo} are affected by the choice of renormalization prescription. 
\acknowledgments{TM would like to thank Esteban Calzetta for helpful discussions and Vincent Vennin for thoroughly reading the manuscript and providing insightful comments towards its improvement. This work is supported by STFC grant ST/P000762/1.}


\begin{thebibliography}{99}

\bibitem{Guth:1980zm}
  A.~H.~Guth,
  ``The Inflationary Universe: A Possible Solution to the Horizon and Flatness Problems,''
  Phys.\ Rev.\ D {\bf 23} (1981) 347.
  doi:10.1103/PhysRevD.23.347
  
\bibitem{Linde:1981mu}
  A.~D.~Linde,
  ``A New Inflationary Universe Scenario: A Possible Solution of the Horizon, Flatness, Homogeneity, Isotropy and Primordial Monopole Problems,''
  Phys.\ Lett.\  {\bf 108B} (1982) 389.
  doi:10.1016/0370-2693(82)91219-9
  
\bibitem{Albrecht:1982wi}
  A.~Albrecht and P.~J.~Steinhardt,
  ``Cosmology for Grand Unified Theories with Radiatively Induced Symmetry Breaking,''
  Phys.\ Rev.\ Lett.\  {\bf 48} (1982) 1220.
  doi:10.1103/PhysRevLett.48.1220
  
\bibitem{Starobinsky:1979ty}
  A.~A.~Starobinsky,
  ``Spectrum of relict gravitational radiation and the early state of the universe,''
  JETP Lett.\  {\bf 30} (1979) 682
   [Pisma Zh.\ Eksp.\ Teor.\ Fiz.\  {\bf 30} (1979) 719].
  
\bibitem{Starobinsky:1980te}
  A.~A.~Starobinsky,
  ``A New Type of Isotropic Cosmological Models Without Singularity,''
  Phys.\ Lett.\  {\bf 91B} (1980) 99.
  doi:10.1016/0370-2693(80)90670-X

\bibitem{Ade:2015lrj}
  P.~A.~R.~Ade {\it et al.} [Planck Collaboration],
  ``Planck 2015 results. XX. Constraints on inflation,''
  Astron.\ Astrophys.\  {\bf 594} (2016) A20
  doi:10.1051/0004-6361/201525898
  [arXiv:1502.02114 [astro-ph.CO]].

\bibitem{Sasaki:1986hm}
  M.~Sasaki,
  ``Large Scale Quantum Fluctuations in the Inflationary Universe,''
  Prog.\ Theor.\ Phys.\  {\bf 76} (1986) 1036.
  doi:10.1143/PTP.76.1036
  
\bibitem{Mukhanov:1988jd}
  V.~F.~Mukhanov,
  ``Quantum Theory of Gauge Invariant Cosmological Perturbations,''
  Sov.\ Phys.\ JETP {\bf 67} (1988) 1297
   [Zh.\ Eksp.\ Teor.\ Fiz.\  {\bf 94N7} (1988) 1].

	\bibitem{Parker:2007}
	  L.~Parker,
	  ``Amplitude of Perturbations from Inflation,''
	  hep-th/0702216 [HEP-TH].

	\bibitem{Agullo:2008ka}
	  I.~Agullo, J.~Navarro-Salas, G.~J.~Olmo and L.~Parker,
	  ``Reexamination of the Power Spectrum in De Sitter Inflation,''
	  Phys.\ Rev.\ Lett.\  {\bf 101} (2008) 171301
	  doi:10.1103/PhysRevLett.101.171301
	  [arXiv:0806.0034 [gr-qc]].
	
	\bibitem{Agullo:2009vq}
	  I.~Agullo, J.~Navarro-Salas, G.~J.~Olmo and L.~Parker,
	  ``Revising the predictions of inflation for the cosmic microwave background anisotropies,''
	  Phys.\ Rev.\ Lett.\  {\bf 103} (2009) 061301
	  doi:10.1103/PhysRevLett.103.061301
	  [arXiv:0901.0439 [astro-ph.CO]].
	  
	\bibitem{Agullo:2009zza}
	  I.~Agullo, J.~Navarro-Salas, G.~J.~Olmo and L.~Parker,
	  ``Inflation, quantum fields, and CMB anisotropies,''
	  Gen.\ Rel.\ Grav.\  {\bf 41} (2009) 2301
	   [Int.\ J.\ Mod.\ Phys.\ D {\bf 18} (2009) 2329]
	  doi:10.1142/S0218271809016144, 10.1007/s10714-009-0850-6
	  [arXiv:0909.0026 [gr-qc]].
	
	\bibitem{Agullo:2009zi}
	  I.~Agullo, J.~Navarro-Salas, G.~J.~Olmo and L.~Parker,
	  ``Revising the observable consequences of slow-roll inflation,''
	  Phys.\ Rev.\ D {\bf 81} (2010) 043514
	  doi:10.1103/PhysRevD.81.043514
	  [arXiv:0911.0961 [hep-th]].
	  
	\bibitem{Agullo:2010ui}
	  I.~Agullo, J.~Navarro-Salas, G.~J.~Olmo and L.~Parker,
	  ``Inflation, Renormalization, and CMB Anisotropies,''
	  J.\ Phys.\ Conf.\ Ser.\  {\bf 229} (2010) 012058
	  doi:10.1088/1742-6596/229/1/012058
	  [arXiv:1002.3914 [gr-qc]].
	  
	\bibitem{Agullo:2010hg}
	  I.~Agullo, J.~Navarro-Salas, G.~J.~Olmo and L.~Parker,
	  ``Inflation, Quantum Field Renormalization, and CMB Anisotropies,''
	  doi:10.1142/9789814374552-0174
	  arXiv:1005.2727 [astro-ph.CO].
	  
	\bibitem{Agullo:2011qg}
	  I.~Agullo, J.~Navarro-Salas, G.~J.~Olmo and L.~Parker,
	  ``Remarks on the renormalization of primordial cosmological perturbations,''
	  Phys.\ Rev.\ D {\bf 84} (2011) 107304
	  doi:10.1103/PhysRevD.84.107304
	  [arXiv:1108.0949 [gr-qc]].
  
	\bibitem{Bastero-Gil:2013}
	  M.~Bastero-Gil, A.~Berera, N.~Mahajan and R.~Rangarajan,
	  ``Power spectrum generated during inflation,''
	  Phys.\ Rev.\ D {\bf 87} (2013) 8,  087302
	  [arXiv:1302.2995 [astro-ph.CO]].	
	  
	\bibitem{Durrer:2009ii}
	  R.~Durrer, G.~Marozzi and M.~Rinaldi,
	  ``On Adiabatic Renormalization of Inflationary Perturbations,''
	  Phys.\ Rev.\ D {\bf 80} (2009) 065024
	  [arXiv:0906.4772 [astro-ph.CO]].	  
	  
	\bibitem{Marozzi:2011da}
	  G.~Marozzi, M.~Rinaldi and R.~Durrer,
	  ``On infrared and ultraviolet divergences of cosmological perturbations,''
	  Phys.\ Rev.\ D {\bf 83} (2011) 105017
	  doi:10.1103/PhysRevD.83.105017
	  [arXiv:1102.2206 [astro-ph.CO]].		  

	\bibitem{Finelli:2007fr}
	  F.~Finelli, G.~Marozzi, G.~P.~Vacca and G.~Venturi,
	  ``The Impact of ultraviolet regularization on the spectrum of curvature perturbations during inflation,''
	  Phys.\ Rev.\ D {\bf 76} (2007) 103528
	  [arXiv:0707.1416 [hep-th]].

\bibitem{Glenz:2009zn}
  M.~M.~Glenz and L.~Parker,
 ``Study of the Spectrum of Inflaton Perturbations,''
  Phys.\ Rev.\ D {\bf 80} (2009) 063534
  doi:10.1103/PhysRevD.80.063534
  [arXiv:0905.2624 [hep-th]].
  
\bibitem{Urakawa:2009xaa}
  Y.~Urakawa and A.~A.~Starobinsky,
  ``Adiabatic regularization of primordial perturbations generated during inflation,''
  
\bibitem{Seery:2010kh}
  D.~Seery,
  ``Infrared effects in inflationary correlation functions,''
  Class.\ Quant.\ Grav.\  {\bf 27} (2010) 124005
  doi:10.1088/0264-9381/27/12/124005
  [arXiv:1005.1649 [astro-ph.CO]].
	  
\bibitem{Woodard:2014jba}
  R.~P.~Woodard,
  ``Perturbative Quantum Gravity Comes of Age,''
  Int.\ J.\ Mod.\ Phys.\ D {\bf 23} (2014) no.09,  1430020
  doi:10.1142/S0218271814300201
  [arXiv:1407.4748 [gr-qc]].
  
\bibitem{Alinea:2015pza}
  A.~L.~Alinea, T.~Kubota, Y.~Nakanishi and W.~Naylor,
  ``Adiabatic regularisation of power spectra in $k$-inflation,''
  JCAP {\bf 1506} (2015) no.06,  019
  doi:10.1088/1475-7516/2015/06/019
  [arXiv:1503.08073 [gr-qc]].
  
\bibitem{Alinea:2016qlf}
  A.~L.~Alinea,
  ``Adiabatic regularization of power spectra in nonminimally coupled chaotic inflation,''
  JCAP {\bf 1610} (2016) no.10,  027
  doi:10.1088/1475-7516/2016/10/027
  [arXiv:1607.05328 [gr-qc]].

\bibitem{Alinea:2017ncx}
  A.~L.~Alinea,
  ``Adiabatic regularisation of power spectrum in nonminimally coupled general single-field inflation,''
  arXiv:1709.06450 [gr-qc].


\bibitem{Parker:1974qw}
  L.~Parker and S.~A.~Fulling,
  ``Adiabatic regularization of the energy momentum tensor of a quantized field in homogeneous spaces,''
  Phys.\ Rev.\ D {\bf 9} (1974) 341.
  
\bibitem{Fulling:1974zr}
  S.~A.~Fulling and L.~Parker,
  ``Renormalization in the theory of a quantized scalar field interacting with a robertson-walker spacetime,''  Annals Phys.\  {\bf 87} (1974) 176.

  \bibitem{Parker:1974qw1}
 S.~A.~Fulling, L.~Parker and B.~L.~Hu,
  ``Conformal energy-momentum tensor in curved spacetime: Adiabatic regularization and renormalization,''
   Phys.\ Rev.\ D {\bf 10} (1974) 3905.

\bibitem{Bunch:1980vc}
  T.~S.~Bunch,
  ``Adiabatic Regularization For Scalar Fields With Arbitrary Coupling To The Scalar Curvature,''
  J.\ Phys.\ A {\bf 13} (1980) 1297.  
  
\bibitem{Anderson:1987yt}
  P.~R.~Anderson and L.~Parker,
  ``Adiabatic Regularization in Closed Robertson-walker Universes,''
  Phys.\ Rev.\ D {\bf 36} (1987) 2963.
  doi:10.1103/PhysRevD.36.2963
  
\bibitem{Habib:1999cs}
  S.~Habib, C.~Molina-Paris and E.~Mottola,
  ``Energy momentum tensor of particles created in an expanding universe,''
  Phys.\ Rev.\ D {\bf 61} (2000) 024010
  doi:10.1103/PhysRevD.61.024010
  [gr-qc/9906120].
  
\bibitem{MolinaParis:2000zz}
  C.~Molina-Paris, P.~R.~Anderson and S.~A.~Ramsey,
  ``One-loop lamdaphi4 field theory in Robertson-Walker spacetimes: Adiabatic regularization and analytic approximations,''
  Phys.\ Rev.\ D {\bf 61} (2000) 127501.
  doi:10.1103/PhysRevD.61.127501
 
\bibitem{Brandenberger:2004kx}
  R.~H.~Brandenberger and J.~Martin,
  ``Back-reaction and the trans-Planckian problem of inflation revisited,''
  Phys.\ Rev.\ D {\bf 71} (2005) 023504
  doi:10.1103/PhysRevD.71.023504
  [hep-th/0410223].
  
\bibitem{Anderson:2005hi}
  P.~R.~Anderson, C.~Molina-Paris and E.~Mottola,
  ``Short distance and initial state effects in inflation: Stress tensor and decoherence,''
  Phys.\ Rev.\ D {\bf 72} (2005) 043515
  doi:10.1103/PhysRevD.72.043515
  [hep-th/0504134].
  
\bibitem{Kaya:2015hka}
  A.~Kaya and E.~S.~Kutluk,
  ``Adiabatic regularization of functional determinants in cosmology and radiative corrections during inflation,''
  Phys.\ Rev.\ D {\bf 92} (2015) 123505
  doi:10.1103/PhysRevD.92.123505
  [arXiv:1509.00489 [gr-qc]].

    
     
\bibitem{Collins:1984xc}
  J.~C.~Collins,
  ``Renormalization : An Introduction to Renormalization, The Renormalization Group, and the Operator Product Expansion,''
  Cambridge University Press.
   

\bibitem{Markkanen:2013nwa}
  T.~Markkanen and A.~Tranberg,
  ``A Simple Method for One-Loop Renormalization in Curved Space-Time,''
  JCAP {\bf 1308} (2013) 045
  doi:10.1088/1475-7516/2013/08/045
  [arXiv:1303.0180 [hep-th]].

\bibitem{Cooper:1994ji}
  F.~Cooper, Y.~Kluger, E.~Mottola and J.~P.~Paz,
  ``Nonequilibrium quantum dynamics of disoriented chiral condensates,''
  Phys.\ Rev.\ D {\bf 51} (1995) 2377
  doi:10.1103/PhysRevD.51.2377
  [hep-ph/9404357].

\bibitem{Cooper:1996ib}
  F.~Cooper, Y.~Kluger and E.~Mottola,
  ``Anomalous transverse distribution of pions as a signal for the production of DCCs,''
  Phys.\ Rev.\ C {\bf 54} (1996) 3298
  doi:10.1103/PhysRevC.54.3298
  [hep-ph/9604284].

\bibitem{Lampert:1996qw}
  M.~A.~Lampert, J.~F.~Dawson and F.~Cooper,
  ``Time evolution of the chiral phase transition during a spherical expansion,''
  Phys.\ Rev.\ D {\bf 54} (1996) 2213
  doi:10.1103/PhysRevD.54.2213
  [hep-th/9603068].

\bibitem{Calzetta:2008iqa}
  E.~A.~Calzetta and B.~L.~B.~Hu,
  ``Nonequilibrium Quantum Field Theory,''
  Cambridge University Press (2008)

\bibitem{Granese:2017gfb}
  N.~Miron-Granese and E.~Calzetta,
  arXiv:1709.01661 [gr-qc].


\bibitem{Birrell:1982ix}
  N.~D.~Birrell and P.~C.~W.~Davies,
  ``Quantum Fields in Curved Space,''
  Cambridge University Press (1982).

\bibitem{Parker:2009uva}
  L.~E.~Parker and D.~Toms,
  ``Quantum Field Theory in Curved Spacetime : Quantized Field and Gravity,''
  doi:10.1017/CBO9780511813924
  
\bibitem{Misner:1974qy}
  C.~W.~Misner, K.~S.~Thorne and J.~A.~Wheeler,
  ``Gravitation,''
  San Francisco 1973, W. H. Freeman and Company, p. 1279.
    
\bibitem{Chernikov:1968zm}
  N.~A.~Chernikov and E.~A.~Tagirov,
  ``Quantum theory of scalar fields in de Sitter space-time,''
  Annales Poincare Phys.\ Theor.\ A {\bf 9} (1968) 109.
\bibitem{BD} 
  T.~S.~Bunch and P.~C.~W.~Davies,
  ``Quantum Field Theory in de Sitter Space: Renormalization by Point Splitting,''
  Proc.\ Roy.\ Soc.\ Lond.\ A {\bf 360} (1978) 117.

\bibitem{Peskin:1995ev}
  M.~E.~Peskin and D.~V.~Schroeder,
  ``An Introduction to quantum field theory,''
  
\bibitem{Wald:1977up}
  R.~M.~Wald,
  Commun.\ Math.\ Phys.\  {\bf 54} (1977) 1.
  doi:10.1007/BF01609833
  
\bibitem{Wald:1978pj}
  R.~M.~Wald,
  Phys.\ Rev.\ D {\bf 17} (1978) 1477.
  doi:10.1103/PhysRevD.17.1477
  
\bibitem{Wald:1978ce}
  R.~M.~Wald,
  Annals Phys.\  {\bf 110} (1978) 472.
  doi:10.1016/0003-4916(78)90040-4

\bibitem{Maggiore:2010wr2}  
  L.~Hollenstein, M.~Jaccard, M.~Maggiore and E.~Mitsou,
  ``Zero-point quantum fluctuations in cosmology,''
  Phys.\ Rev.\ D {\bf 85} (2012) 124031
  doi:10.1103/PhysRevD.85.124031
  [arXiv:1111.5575 [astro-ph.CO]].
  
\bibitem{Akhmedov:2002ts}
E.~K.~Akhmedov,
``Vacuum energy and relativistic invariance,''
hep-th/0204048.
  
\bibitem{Markkanen:2016aes}
  T.~Markkanen and A.~Rajantie,
  ``Massive scalar field evolution in de Sitter,''
  JHEP {\bf 1701} (2017) 133
  doi:10.1007/JHEP01(2017)133
  [arXiv:1607.00334 [gr-qc]].

\bibitem{Starobinsky:1986fx}
  A.~A.~Starobinsky,
  ``Stochastic De Sitter (inflationary) Stage In The Early Universe,''
  Lect.\ Notes Phys.\  {\bf 246} (1986) 107.
  doi:10.1007/3-540-16452-9 6

\bibitem{Starobinsky:1994bd}
  A.~A.~Starobinsky and J.~Yokoyama,
  ``Equilibrium state of a selfinteracting scalar field in the De Sitter background,''
  Phys.\ Rev.\ D {\bf 50} (1994) 6357
  doi:10.1103/PhysRevD.50.6357
  [astro-ph/9407016].
  
\bibitem{Grain:2017dqa}
  J.~Grain and V.~Vennin,
  ``Stochastic inflation in phase space: Is slow roll a stochastic attractor?,''
  JCAP {\bf 1705} (2017) no.05,  045
  doi:10.1088/1475-7516/2017/05/045
  [arXiv:1703.00447 [gr-qc]].
  
\bibitem{Hardwick:2017fjo}
  R.~J.~Hardwick, V.~Vennin, C.~T.~Byrnes, J.~Torrado and D.~Wands,
  ``The stochastic spectator,''
  JCAP {\bf 1710} (2017) 018
  doi:10.1088/1475-7516/2017/10/018
  [arXiv:1701.06473 [astro-ph.CO]].
  
\end{thebibliography}
\end{document}